\documentclass[reqno,12pt]{article}
\pdfoutput=1

\usepackage[english]{babel}
\usepackage{ae}
\usepackage[T1]{fontenc}
\usepackage{upgreek}
\usepackage[ansinew]{inputenc}
\usepackage{mathrsfs}
\usepackage{amsmath,amsfonts,amssymb,amsbsy,dsfont}
\usepackage{graphicx,color}
\usepackage{graphics}
\usepackage{verbatim}
\usepackage{stmaryrd}
\usepackage{ulem}
\usepackage{wrapfig}
\usepackage{hyperref}

\def\bc{\begin{center}}
\def\ec{\end{center}}

\definecolor{darkblue}{cmyk}{0.9,0.9,0,0}
\definecolor{darkgreen}{rgb}{0,0.55,0}

\newcommand{\be}{\begin{equation}}
\newcommand{\ee}{\end{equation}}

\newcommand{\ba}{\begin{eqnarray}}
\newcommand{\ea}{\end{eqnarray}}

\newcommand{\nn}{{\nonumber}}

\definecolor{darkgreen}{rgb}{0.1,0.7,0.1}

\begin{document}

\thispagestyle{empty}

\renewcommand{\thefootnote}{\fnsymbol{footnote}}
\setcounter{footnote}{0}
\setcounter{figure}{0}
\begin{center}
$$$$
{\Large
\textbf{D5-brane boundary reflection factors}\par}

\vspace{1.0cm}

\textrm{Diego H. Correa and  Fidel I. Schaposnik Massolo}
\\ \vspace{1.2cm}
\footnotesize{

\textit{Instituto de F\'{\i}sica La Plata, CONICET
\\
Departamento de F\'isica, Facultad de Ciencias Exactas, Universidad Nacional de La Plata
\\
 C.C. 67, 1900 La Plata, Argentina} \\
\texttt{} \\
\vspace{3mm}
}

\par\vspace{1.5cm}

\textbf{Abstract}\vspace{2mm}
\end{center}

\noindent

We compute the strong coupling limit of the boundary reflection factor for excitations
on open strings attached to various kinds of D5-branes that probe AdS$_5\times$S$^5$.
We study the crossing equation, which constrains the boundary reflection factor,
and propose that some solutions will give the boundary reflection factors for
all values of the coupling. Our proposal passes various checks in the strong
coupling limit by comparison with diverse explicit string theory computations.
In some of the cases we consider, the D5-branes correspond to $\tfrac{1}{2}$-BPS Wilson loops
in the $k$-th rank antisymmetric representation of the dual field theory. In the other cases
they correspond in the dual field theory to the addition of a fundamental hypermultiplet in a defect.

\vspace*{\fill}

\setcounter{page}{1}
\renewcommand{\thefootnote}{\arabic{footnote}}
\setcounter{footnote}{0}

\newpage

\tableofcontents

\section{Introduction}

Due to the underlying integrability in its planar limit, ${\cal N}=4$ super Yang-Mills is the better understood interacting four-dimensional
non-abelian gauge theory (see the review \cite{Beisert:2010jr} and references therein). In the strong coupling limit, the integrability is that
of the two-dimensional field theory defined on the worldsheet of the dual string that propagates in AdS$_5\times$S$^5$.

Integrable two-dimensional systems can also be formulated in a half-line if suitable boundary conditions preserving integrability are
imposed. Then, it is reasonable to enquire about the integrability of open strings in the background of AdS$_5\times$S$^5$. The classical
integrability of open strings attached to various kinds of D-branes has been analyzed in \cite{Mann:2006rh,Dekel:2011ja}. In many of those
situations, the symmetries of the problem are enough to fix the boundary scattering matrix exactly, up to an overall reflection factor, as a
function of the coupling \cite{Hofman:2007xp,Correa:2008av,Correa:2011nz}. In all these cases the resulting reflection matrix was
shown to satisfy the boundary Yang-Baxter condition. Determining the remaining overall reflection factor exactly is the last step missing to obtain
an exact description by means of Bethe ansatz techniques. As usual, this overall factor can be constrained by the imposition of crossing symmetry.
However, there are infinitely many different ways of solving this boundary crossing condition. Thus, having explicit computations for the
reflection factor in some limits is indispensable for picking the right solution to the crossing equation.

In this article we compute the boundary reflection factor in the strong coupling limit for excitations propagating
along open strings with large angular momentum attached to certain kinds of D5-branes, and study solutions of the crossing equations
consistent with them. More specifically, we consider two families of D5-branes in the background of AdS$_5\times$S$^5$. The first family contains D5-branes whose worldvolume has the geometry of AdS$_2\times$S$^4$ and an electric field in the AdS$_2$ factor.
The second family contains D5-branes whose worldvolume has the geometry of AdS$_4\times$S$^2$ and a magnetic
field in the S$^2$.

All these D5-branes are $\tfrac{1}{2}$-BPS and the two families have different interpretations in the dual conformal field theory. The D5-branes of the first family are the dual description of  $\tfrac{1}{2}$-BPS Wilson loops in the $k$-th rank antisymmetric representation of the SU($N$) in ${\cal N}=4$ super Yang-Mills theory \cite{Yamaguchi:2006tq}, where $k$ is related to the amount of electric flux in the D5-brane.
Actually, the relation between certain D5-branes and multi-quark states had already been
pointed out in \cite{Callan:1998iq}. The D5-branes we consider here in the first family
are a limiting case of those other ones \cite{Chernicoff:2006yp}.
The matrix structure in the corresponding scattering problem is fixed by the underlying symmetry, which is in this case a diagonal $su(2|2)$ of the usual $su(2|2)^2$ for the case with no boundaries. Certainly, the underlying symmetry is independent of $k$, so for all values of $k$ the matrix structure of the reflection is same. In the limiting case of $k=1$, for which the size of its $\rm{S}^4$ shrinks to zero and the D5-brane reduces to the string dual to a fundamental $\tfrac{1}{2}$-BPS Wilson loop, this matrix structure has been obtained in \cite{Correa:2012hh,Drukker:2012de}. Thus,
the boundary reflection matrix for the D5-branes in this case is the same as the one for the string dual to the Wilson loop in the fundamental representation \cite{Correa:2012hh,Drukker:2012de}. The difference will be at most in the overall reflection factor, which is not fixed by symmetry arguments.

The D5-branes of the second family are interpreted in the dual conformal field theory as having fundamental hypermultiplets living on a
2+1-dimensional defect in addition to ${\cal N}=4$ super Yang-Mills \cite{DeWolfe:2001pq}. The addition of magnetic flux in the D5-brane is interpreted in the dual defect theory as if some fields of the fundamental hypermultiplet had acquired a vacuum expectation value \cite{Arean:2006vg}. In this case, the underlying symmetry that constrains the reflection matrix is also the same independently of the amount of magnetic flux. Then, the matrix structure of the reflection is the same one found in \cite{Correa:2008av}.

This paper is organized as follows. In section \ref{classical} we present classical open strings carrying large angular momentum along the $\rm{S}^5$ and with their endpoints attached to D5-branes of the sorts discussed above. Then, in section \ref{timedelay} we study  excitations that propagate in the worldsheet and compute the time delays during their reflections, which allow us to obtain the boundary reflection factors in the strong coupling regime. We proceed in section \ref{obliquebranes} to compute the difference between energy and angular momentum for strings attached to a pair of oblique D5-branes, in the limit of large but finite angular momentum. In section \ref{crossingsolutions} we analyze different solutions of the boundary crossing and unitarity equations which are consistent with the results obtained in sections  \ref{timedelay} and \ref{obliquebranes}. We summarize and discuss our results in section \ref{conclusions}.

\section{Classical strings ending on D5-branes with fluxes} \label{classical}

In this section we describe semiclassical open strings rotating in ${\rm AdS}_5 \times {\rm S}^5$, whose endpoints are attached to certain kinds
of D5-branes. In first place, we will consider the case in which they carry a large amount $L$ of five-sphere angular momentum  and
have $E-L = 0$. Later on, we will use these configurations as reference states along which impurities can propagate.

Let us begin by describing the D5-branes we will use to impose boundary conditions to the open strings.
We will analyze two families of D5-branes:
\begin{enumerate}
    \item D5-branes with AdS$_2\times$S$^4$ worldvolume and an electric field;
    \item D5-branes with AdS$_4\times$S$^2$ worldvolume and a magnetic field.
\end{enumerate}

\begin{figure}[h]
\centering
\def\svgwidth{9cm}
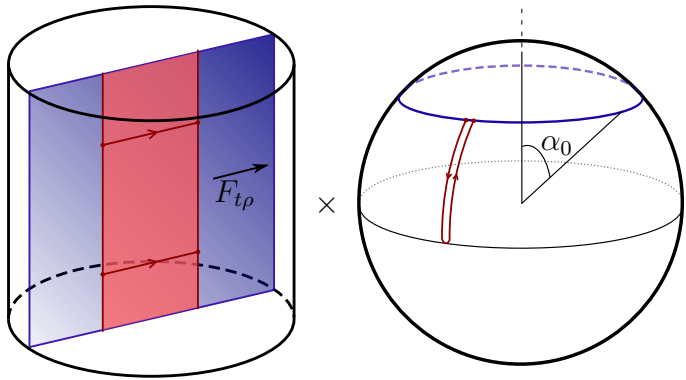
\caption{In blue we depict the D5-brane worldvolume, {\it i.e.} an ${\rm AdS}_2$ factor within ${\rm AdS}_5$ and a $\rm{S}^4$ within the $\rm{S}^5$.
In red we draw the worldsheet of a string with large angular momentum attached to this D5-brane.}
\label{figD5x}
\end{figure}

If we write the metric of AdS$_5\times$S$^5$ in global coordinates
\be\label{global}
ds^2 = R^2(-\cosh^2\rho \, dt ^2 + d\rho^2 + \sinh^2\rho \, d\Omega_3^2 +d\alpha^2+\sin^2\alpha \, d\Omega_4^2)\,,
\ee
the D5-branes of the first family are extended along $t , \rho$ and $\Omega_4$, while they sit at a fixed
value $\alpha_0$ of the azimuthal angle. This value is related to the intensity of the electric
field in the AdS factor of the worldvolume by\footnote{In the conventions we follow $g=\frac{R^2}{4\pi\alpha'} = \frac{\sqrt{\lambda}}{4\pi}$.}
\be\label{solu02}
F = F_{t \rho} \, dt \wedge d\rho\,, \qquad \text{with} \qquad F_{t \rho} = \pm 2g \cosh\rho \cos\alpha_0\,.
\ee
Half of this worldvolume is at some point\footnote{For the 3-sphere in AdS we use
\be
d\Omega_3^2 = d\psi_1^2 + \sin^2\psi_1 \left(d\psi_2^2 + \sin^2\psi_2 d\beta^2\right)\,.
\nn
\ee
} of the $\Omega_3$ sitting in AdS specified by $\beta = \beta_0$ and $\psi_1=\psi_2=\frac\pi2$. The other half is at $\beta_0+\pi$ and $\psi_1=\psi_2=\frac\pi2$ (see figure \ref{figD5x}). Then the $\pm$ signs above correspond to the sheets at $\beta=\beta_0$ and $\beta=\beta_0 +\pi$ respectively.

When there is no electric field, the S$^4$ of the worldvolume is of maximal size and sits on the equator
of the S$^5$. On the other hand, when electric flux is turned on in the D5-brane, the S$^4$ of the worldvolume is displaced away
from the equator. The amount of electric flux is discretized according to \cite{Pawelczyk:2000hy,Camino:2001at}
\be
\frac{k}{N} = \frac{\alpha_0}{\pi} -\frac{\sin2\alpha_0}{2\pi}\,,
\ee
where $k$ is an integer. As said before, these D5-branes
are dual to BPS Wilson loops in the antisymmetric representation
and the integer $k$ is in correspondence with the rank of this representation \cite{Yamaguchi:2006tq}.

\begin{figure}[h]
\centering
\def\svgwidth{9cm}
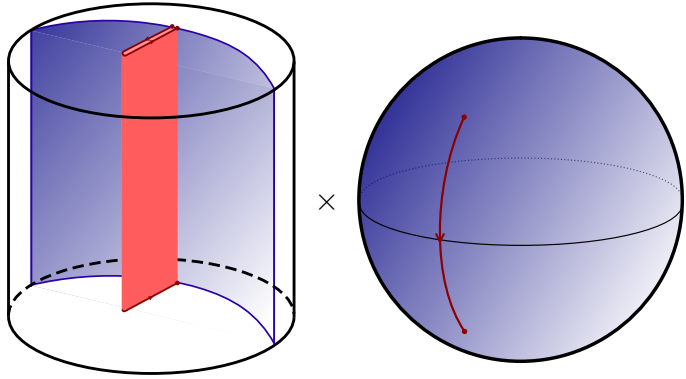
\caption{In blue we depict the D5-brane worldvolume, {\it i.e.} an ${\rm AdS}_4$ factor within ${\rm AdS}_5$ and a $\rm{S}^2$.
In red we draw the worldsheet of a string with large angular momentum attached to this D5-brane.}
\label{figD52}
\end{figure}

The second family of D5-brane solutions has been found in \cite{Karch:2000gx}.
In this case the AdS factor of the worldvolume is defined through the radial position
of the brane as a function of the angular position in the S$^3\subset {\rm AdS}_5$, as schematically depicted in figure \ref{figD52}.
In the S$^5$ the D5-brane is extended along the azimuthal angle $\alpha$ and a circle in $\Omega_4$.
This defines an S$^2$ on which a magnetic field can be turned on,
\be
F = F_{\alpha \varphi} \, d\alpha \wedge d\varphi = \frac{q}2 \sin\alpha \, d\alpha \wedge d\varphi\,,
\ee
where $q$ is the integer that specifies magnetic flux. The D5-brane probes the interior
of AdS from the boundary to a distance $\rho_0$  given by
\be
\sinh\rho_0 = \frac{|q|}{4g}\,.
\ee

\subsection{Semiclassical strings}
\label{semicla}

In what follows we will present open string solutions carrying a large amount of angular momentum $L$ and attached to D5-branes of the sort described above. For the time being, we only consider folded string solutions, extended along the azimuthal angle
of the sphere and the radial coordinate of AdS. More general configurations will be studied later on.
Thus, we will look for strings extended along directions $\rho$ and $\alpha$, while spinning around some
$\varphi$  which parametrizes a circle in the $\rm{S}^4$,
\begin{alignat}{4}
t &= \tau\,, & \qquad \quad &  & \varphi&=  \omega\tau\,, \\
\rho &= \rho(\sigma)\,, & \qquad \quad & & \alpha&=  \alpha(\sigma)\,.
\end{alignat}
The equations of motion can be obtained from the Nambu-Goto action,
\be\label{reducedNG}
S_{\rm NG} = - 2g \int d^2\sigma\sqrt{(\cosh^2\rho-\omega^2\sin^2\alpha)({\rho'}^2+{\alpha'}^2)}\,.
\ee
This action should be supplemented with boundary terms
\be\label{bterm}
S_{\rm bdry} = \int d\tau A_\mu \left.\frac{dX^\mu}{d{\tau}}\right|_{\sigma=\pi}- \int d\tau A_\mu \left.\frac{dX^\mu}{d{\tau}}\right|_{\sigma={0}}\,,
\ee
when the D5-branes carry electromagnetic fields.

Let us focus on the boundary conditions set by a D5-brane of the first family.
For  $\alpha(\sigma)$ they are of Dirichlet type, and the endpoints of the string
are then forced to be at $\alpha_0$. Because of the angular momentum, the string will be stretched
away from $\alpha_0$ towards the equator of the S$^5$. In the limit $L\to\infty$ the folded string
will be extended from $\alpha_0$ all the way to the equator and back to $\alpha_0$.
In AdS the string will be stretched from $0$ to some $\rho_0$ because the electric field will pull its endpoints.
From now on, we will concentrate on one half of the folded string so the
boundary condition driven by a term like \eqref{bterm} will apply to the right endpoint only,
while the left endpoint will be moving along a null geodesic, {\it i.e.} $\rho=0$ and $\alpha=\tfrac{\pi}{2}$.

The string can be parametrized by $\alpha$ and it is then easy to check that the equations of motion
are solved with
\be
\cosh\rho =\frac{1}{\sin\alpha} \qquad \text{and} \qquad  \omega = 1.
\label{solu}
\ee
Concerning the boundary condition for the right endpoint, we have
\be
\left.\left(\frac{\partial {\cal L}}{\partial \rho'} + F_{t \rho}\right)\right|_{\alpha=\alpha_0} = 0\,.
\ee
It is straightforward to verify that the solution \eqref{solu} satisfies this condition\footnote{In the conventions we are using the right endpoint sits at $\beta_0+\pi$.},
\be
\left.\left(\frac{\partial {\cal L}}{\partial \rho'} + F_{t \rho}\right)\right|_{\alpha=\alpha_0} = 2g \left(\cot\alpha_0 - \cos\alpha_0\cosh\rho(\alpha_0)\right) =0\,.
\ee

This solution is a {\it fraction} of the one found by Drukker and Kawamoto
in \cite{Drukker:2006xg}, and reduces to it in the limit of $\alpha_0 \to 0$.

~

Now we would like to compute the energy and the angular momentum of this solution.
Both $E$ and $L$ are divergent, but we are actually interested in the difference $E-L$.
There are  two kinds of contributions to the difference, from the bulk density and from the boundary term,
and they cancel exactly,
\begin{align}
E-L &= 2g \int_{\pi/2}^{\alpha_0} d\alpha \frac{1}{\cos\alpha }\left(\frac{1}{\sin^2\alpha } - \sin^2\alpha\right) + \left.\vphantom{\int} A_t  \right|_{\alpha=\alpha_0}\nn
\\
&= -2g \frac{\cos^2\alpha_0}{\sin\alpha_0} + 2g \cos\alpha_0\sinh\rho(\alpha_0) = 0\,.
\end{align}
Here, we used
\be
A_{t } = 2g \sinh\rho \cos\alpha_0\,,
\ee
which is the gauge potential leading to \eqref{solu02} for the right endpoint.

~

In this parametrization, the density of angular momentum becomes infinite as $\alpha$ approaches $\tfrac{\pi}{2}$.
Alternatively, we can parametrize the same solution in terms of a
semi-infinite spatial coordinate  $x \in (-\infty,0]$:
\begin{align}
& \rho = {\rm arccosh}\left(\tfrac{1}{\tanh (x_0-x)}\right)\,,
\label{xin1}
\\
& \alpha = \arccos\left(\tfrac{1}{\cosh (x_0-x)}\right)\,,
\label{xin2}
\\
& \varphi = t\,,
\label{xin3}
\end{align}
where $\cosh x_0 =\tfrac{1}{\cos\alpha_0}$ and $x_0 > 0$. In this gauge the solution is a static soliton in a semi-infinite line. Far from the soliton, {\it i.e.} for $x \ll 0$, the density of angular momentum becomes constant.

~

Let us now turn our attention to open strings ending on D5-branes of the second family. The boundary terms will be different, leading to different boundary conditions. In this case, it is more natural to use $\rho$ to parametrize the string with $0 < \rho < \rho_0$, and we will then have the right endpoint fixed at $\rho_0$. The boundary condition for the right endpoint is now
\be
\left.\left(\frac{\partial \mathcal{L}}{\partial \alpha'} + F_{\varphi\alpha}\right)\right|_{\rho = \rho_0} = 0\,.
\ee
Of course, \eqref{solu} is still a solution to the equations of motion. Interestingly, it also satisfies
this other boundary condition and the configuration continues to have $E=L$.

\section{Reflection factor in the strong coupling limit}
\label{timedelay}

We are now going to consider more general classical string solutions.
On top of the static soliton we found in section \ref{classical},
we can add propagating solitons which are reflected off the right boundary.
From the solution that corresponds to a reflecting soliton, we will calculate
the time delay experienced during the reflection and from it we will compute
the reflection phase factor.

By means of a Pohlmeyer reduction, one typically relates classical solutions in a $\rm{S}^2$ \mbox{$\sigma$-model}
to classical solutions in a sine Gordon model \cite{Pohlmeyer:1975nb}. The Pohlmeyer reduction can be generalized to relate solutions in
an $\rm{AdS}_2\times\rm{S}^2$ $\sigma$-model to solutions in a sine/sinh Gordon  model \cite{Grigoriev:2007bu}.
If the $\sigma$-model is defined in the half-line then so will be the sine/sinh Gordon system. We can parametrize the AdS$_2$ and the $\rm{S}^2$ with
\begin{alignat}{4}
\eta^1& = \cosh\rho\cos\tau  \,, & \qquad \quad & & n^1 &=  \sin\alpha\cos\varphi\,, \nn
\\
\eta^2& =  \cosh\rho\sin\tau \,,    &\qquad \quad  & & n^2 &= \sin\alpha\sin\varphi \,,
\\
\eta^3& = \sinh\rho  \,,  & \qquad \quad & &   n^3 &= \cos\alpha\,, \nn
\end{alignat}
where $\eta^i$ and $n^i$ satisfy ${\boldsymbol\eta}\cdot{\boldsymbol\eta} = -(\eta^1)^2 -(\eta^2)^2+(\eta^3)^2 = -1$ and
${\bf n}\cdot{\bf n} = (n^1)^2 +(n^2)^2+(n^3)^2 = 1$. The Virasoro constraints for a string in this parametrization are
\begin{alignat}{4}
\dot{\boldsymbol\eta}^2 + {{\boldsymbol\eta}'}^2 &= -1\,, & \qquad \quad & & \dot{\boldsymbol\eta}\cdot{\boldsymbol\eta}' &= 0\,, \nn
\\
\dot{{\bf n}}^2 + {{\bf n}'}^2& = 1\,,  & \qquad \quad & &
\dot{{\bf n}}\cdot{{\bf n}}' &= 0\,, \nn
\end{alignat}
where ${\boldsymbol\eta}$ or ${\bf n}$ scalar products should be used in each case.

Following the Pohlmeyer reduction, the $\sigma$-model fields are related to a
sine Gordon field ${\boldsymbol \upphi}$ and a sinh Gordon field ${\boldsymbol \upvarphi}$
according to
\begin{align}
\dot{\boldsymbol\eta}^2 -  {{\boldsymbol \eta}'}^2& = - \cosh 2{\boldsymbol\upvarphi}\,,
\\
\dot{{\bf n}}^2 - {{\bf n}'}^2& =  \cos 2{\boldsymbol\upphi}\,.
\end{align}

Let us concentrate on the sine Gordon part of the system. Its equation of motion is
\be
{\boldsymbol\upphi}'' -\ddot{{\boldsymbol\upphi}} =\frac{1}{2}\sin 2{\boldsymbol\upphi}\,.
\ee
In a half-line $x\leq 0$, the most general boundary condition consistent with integrability
is \cite{Ghoshal:1993tm}
\be
\left.{\boldsymbol\upphi}'\right|_{x=0}  = M \left.\sin({\boldsymbol\upphi}-{\boldsymbol\upphi}_0)\right|_{x=0}\,,
\label{integbc}
\ee
where $M$ and ${\boldsymbol\upphi}_0$ are constants. We will now show that the boundary conditions
inherited from the $\sigma$-model with different sorts of D5-brane boundary conditions lie within this class.

\subsection{AdS$_2 \times$S$^4$ D5-brane with electric field}
\label{subsec:delta_ads2s4}

In this case the D5-brane is placed at some value $\alpha_0$, so the
$\sigma$-model fields $\alpha$ and $\varphi$ satisfy Dirichlet
and Neumann boundary conditions respectively,
\be
\left.\dot\alpha\right|_{x=0} = 0\,, \qquad \left.\varphi'\right|_{x=0} = 0\,.
\ee
Thus, the first  Virasoro constraint at the boundary reads
\be
\left.{\alpha'}^2\right|_{x=0} +\sin^2\alpha_0 \left.\dot\varphi^2\right|_{x=0} =1\,.
\ee

The sine Gordon field is related to the $\sigma$-model fields according to,
\be
 \cos 2{\boldsymbol\upphi} = \dot\alpha^2 -{\alpha'}^2 +\sin^2\alpha \left(\dot\varphi^2 -{\varphi'}^2\right)\,,
\label{sinefield}
\ee
and then we conclude that
\be
\left.\sin {\boldsymbol\upphi} \right|_{x=0}  =  \left.{\alpha'}\right|_{x=0}\,,
\qquad
\left.\cos {\boldsymbol\upphi} \right|_{x=0}  =  \sin\alpha_0 \left.\dot\varphi\right|_{x=0}\,.
\ee
By considering the derivative of equation (\ref{sinefield}) and the first Virasoro constraint, we obtain
\be
\left.{\boldsymbol\upphi}'\right|_{x=0}  = -\cot\alpha_0 \left.\cos{\boldsymbol\upphi}\right|_{x=0}\,,
\label{bc}
\ee
which is a boundary condition consistent with integrability, namely of the form (\ref{integbc})
for $M=\cot\alpha_0$ and ${\boldsymbol\upphi}_0 =  \frac\pi{2}$.

The static soliton configuration  (\ref{xin1})-(\ref{xin3}) is a particular
solution satisfying the boundary condition (\ref{bc}). We will now consider more general
solutions. Multisoliton solutions in the sine Gordon model with integrable boundaries are
known \cite{Saleur:1994yh}. To get a travelling soliton that is reflected off the boundary,
one can consider two solitons on the full line $(-\infty,\infty)$, one with velocity $v$ and its
image with respect to $x=0$ with velocity $-v$. For the sort of boundary we are considering, there is also
a soliton at the boundary, so we will consider a static third soliton. For this kind of solutions
satisfying the boundary conditions (\ref{integbc}), the  classical phase shift $a$
is known (cf. (2.15) in \cite{Saleur:1994yh}).  The classical time delay is obtained from it
through the classical relation $\Delta T = a \tfrac{\sqrt{1-v^2}}{v}$, where $v$ is the velocity
of the travelling soliton. As a function of the rapidity $v=\tanh\theta$ the time delay is
\be
\Delta T=\frac{1}{\sinh{\theta}} \log  \left[\pm\tanh^2\tfrac{\theta}{2}\tanh^2\theta \frac{\tanh\tfrac{1}{2}(\theta+i\eta)\tanh\tfrac{1}{2}(\theta-i\eta)}
{\tanh\tfrac{1}{2}(\theta+\zeta)\tanh\tfrac{1}{2}(\theta-\zeta)}\right]\,,
\label{tdelay}
\ee
where $\zeta$ and $\eta$ parametrize $M$ and ${\boldsymbol\upphi}_0$ as
\be
M \cos{\boldsymbol\upphi}_0 = \cosh\zeta \cos\eta\,,
\qquad
M \sin{\boldsymbol\upphi}_0 =  \sinh\zeta \sin\eta\,,
\ee
and the rapidity $\theta$ is related to the energy and momentum
of the $\sigma$-model soliton according to
\be
\cosh\theta = \frac{4g}{\epsilon} = \frac{1}{|\sin\tfrac{p}{2}|}\,.
\ee
The signs $\pm$  in (\ref{tdelay}) correspond to the cases $|\theta| \gtrless \zeta$.
We are interested in the particular type of boundary conditions obtained
when $M = \cot\alpha_0$ and ${\boldsymbol\upphi}_0 =  \frac\pi{2}$.
For them, we get
\be
\Delta T=
2\tan\tfrac{p}{2}\log\left(\cos\tfrac{p}{2}\right)+
\tan\tfrac{p}{2}\log\left[\left(\frac{1-\sin\tfrac{p}{2}}{1+\sin\tfrac{p}{2}}\right)\left(
\frac{\sin\alpha_0+\sin\tfrac{p}{2}}{|\sin\alpha_0-\sin\tfrac{p}{2}|}\right)\right].
\label{tdelay2}
\ee
The second term is the delay due to the static soliton at the boundary.
As expected this term is vanishing for $\alpha_0 \to \tfrac{\pi}{2}$ when there is no boundary soliton.

The time delay is related to the reflection phase $\delta$ of a reflection factor $R=e^{i\delta}$ \cite{Jackiw:1975im}.
More precisely,
\be
\frac{d\epsilon}{dp}\Delta T = \frac{d \delta}{dp}\,,
\label{JackiwWoo}
\ee
which allows us to obtain $\delta$ by integration. We will consider here a right
boundary and split $\delta = \delta_0 + \delta_{\rm extra}$. In this splitting
$\delta_0$ is the reflection phase as if the static soliton had $\alpha_0 = 0$. Since
it has already been computed, here we shall focus on the extra reflection phase
$\delta_{\rm extra}$. In general, reflection and scattering phases depend on the gauge
used in the $\sigma$-model. In particular, in a $\sigma$-model gauge such that the density
of momentum is constant\footnote{If we integrated
(\ref{tdelay2})  for $\alpha_0 = 0$ we would get an extra term $8g\cos(\tfrac{p}{2})$
 because we computed $\Delta T$ in a $\sigma$-model gauge for which the density
 of momentum is not uniform. The computation of this $\delta_0$ in a gauge
 with constant momentum density was done in detail in \cite{Hofman:2007xp}.} $\delta_0$ is
\cite{Hofman:2007xp,Correa:2012hh,Drukker:2012de}
\be
\delta_0 = -8g\cos\tfrac{p}{2}\log\left(\cos\tfrac{p}{2}\right)
-4g\cos\tfrac{p}{2}\log\left(\frac{1-\sin\tfrac{p}{2}}{1+\sin\tfrac{p}{2}}\right)\,.
\ee

For $\delta_{\rm extra}$, in a $\sigma$-model gauge where the density of momentum is not constant, we get
\be\label{delta1}
\delta_{\rm extra} = -4g\cos\tfrac{p}{2}\log\left|\frac{\sin\alpha_0+\sin\tfrac{p}{2}}{\sin\alpha_0-\sin\tfrac{p}{2}}\right|+4g\cos\alpha_0
\log\left|\frac{\sin(\tfrac{p}{2}+\alpha_0)}{\sin(\tfrac{p}{2}-\alpha_0)}\right| + 4 g p(\sin\alpha_0 -1)\,.
\ee
In order to translate it into a gauge where the density of momentum is constant, we have to take into account the length of the boundary soliton, as discussed in detail for the bulk scattering phase in \cite{Hofman:2006xt}. Let $\Delta x$ be the interval of the boundary soliton in our gauge and $\Delta x'$ the interval in a gauge where density of momentum is constant. The latter is related to the total momentum $L$ according to $L = 2g \Delta x'$, and then the change in the length of the boundary soliton is
\be\label{lchange}
2g \Delta x -L = 2g  \int_{-\infty}^0 \left(1-\frac{d L}{dx}\right)dx
= 2g \int_{-\infty}^0 dx \cos^2\alpha(x) = 2g (1-\sin\alpha_0)\,,
\ee
where $\alpha(x)$ is given in (\ref{xin2}). Therefore, in the non-uniform momentum gauge,
the last term in (\ref{delta1})  would be compensated by twice this length change\footnote{The open boundary Bethe equations depend on twice the length of the system.}. Thus, in a gauge where the density of momentum is constant the total right reflection phase is
\begin{align}
\delta = & -8g\cos\tfrac{p}{2}\log\left(\cos\tfrac{p}{2}\right)
-4g\cos\tfrac{p}{2}\log\left(\frac{1-\sin\tfrac{p}{2}}{1+\sin\tfrac{p}{2}}\right)\nn
\\
&-4g\cos\tfrac{p}{2}\log\left|\frac{\sin\alpha_0+\sin\tfrac{p}{2}}{\sin\alpha_0-\sin\tfrac{p}{2}}\right|
+4g\cos\alpha_0\log\left|\frac{\sin(\tfrac{p}{2}+\alpha_0)}{\sin(\tfrac{p}{2}-\alpha_0)}\right|\,.
\label{deltatotal}
\end{align}
Notice that in the limit $\alpha_0\to 0$, the second line in (\ref{deltatotal}) vanishes
and we recover the result for a string stretching to the boundary of AdS \cite{Correa:2012hh,Drukker:2012de}.
On the other hand, when $\tfrac{p}{2} = \pm \alpha_0$ the two extra terms in the second line
appear to have logarithmic divergencies if considered separately, but these cancel out to give
a regular reflection phase in the strong coupling limit.

\subsection{AdS$_4 \times$S$^2$ D5-brane with magnetic field}
\label{subsec:delta_ads4s2}

In this other case the D5-brane spans both
angular coordinates $\alpha$ and $\varphi$, so they will satisfy
Neumann-like boundary conditions but modified due to the magnetic field
living in the $\rm{S}^2$. We will have
\be
\left.\alpha'\right|_{x=0} -\frac{q}2 \left.\sin\alpha \dot\varphi\right|_{x=0}= 0\,,
\qquad
\sin\alpha \left.\varphi'\right|_{x=0} + \frac{q}2 \left. \dot\alpha\right|_{x=0}= 0\,,
\ee
where $q$ measures the amount of magnetic flux in the $\rm{S}^2$.

The first  Virasoro constraint at the boundary imposes
\be
\sin^2\alpha \left.\dot\varphi^2\right|_{x=0} = \frac{1}{1+(\tfrac{q}{2})^2} - \left.{\dot\alpha}^2\right|_{x=0}\,,
\ee
from which we get
\be
\left.\cos 2{\boldsymbol\upphi}\right|_{x=0} = \frac{1-(\tfrac{q}{2})^2}{1+(\tfrac{q}{2})^2} \equiv \cos 2{\boldsymbol\upphi}_0\,.
\ee
Then, we have Dirichlet boundary conditions for the sine Gordon field in this case,
which corresponds to $M\to\infty$ in (\ref{integbc}).

The time delay in this case is obtained from (\ref{tdelay}) by taking $\zeta\to\infty$, and we get
\be
\Delta T=
2\tan\tfrac{p}{2}\log\left(\cos\tfrac{p}{2}\right)+
\tan\tfrac{p}{2}\log\left[\left(\frac{1-\sin\tfrac{p}{2}}{1+\sin\tfrac{p}{2}}\right)\left(\frac{1+\cos{\boldsymbol\upphi}_0 \sin\tfrac{p}{2}}{1-\cos{\boldsymbol\upphi}_0 \sin\tfrac{p}{2}}\right)\right].
\label{tdelay3}
\ee
We can split the reflection phase as before $\delta = \delta_0 +\delta_{\rm extra}$, with
\begin{align}
\delta_{\rm extra} = &-4g\cos\tfrac{p}{2}\log\left(\frac{1+\cos{\boldsymbol\upphi}_0 \sin\tfrac{p}{2}}{1-\cos{\boldsymbol\upphi}_0 \sin\tfrac{p}{2}}\right)
- 8 g \tan {\boldsymbol\upphi}_0 \arctan(\sin {\boldsymbol\upphi}_0 \tan \tfrac{p}{2})\nn\\
& +  4 g p\left(\frac{1}{\cos{\boldsymbol\upphi}_0}-1\right)\,.\label{delta41}
\end{align}

The static soliton at the boundary is the same one considered in the previous section, if we identify $\cos{\boldsymbol\upphi}_0$
with $\sin\alpha_0$. The same term (\ref{lchange}) must then be subtracted to express the reflection phase in a gauge where
the density of momentum is constant. We obtain in this case
\begin{align}
\delta = & -8g\cos\tfrac{p}{2}\log\left(\cos\tfrac{p}{2}\right) -4g\cos\tfrac{p}{2}\log\left(\frac{1-\sin\tfrac{p}{2}}{1+\sin\tfrac{p}{2}}\right)\nn
-4g\cos\tfrac{p}{2}\log\left(\frac{1+\cos{\boldsymbol\upphi}_0 \sin\tfrac{p}{2}}{1-\cos{\boldsymbol\upphi}_0 \sin\tfrac{p}{2}}\right)
\\
&
- 8 g \tan {\boldsymbol\upphi}_0 \arctan(\sin {\boldsymbol\upphi}_0 \tan \tfrac{p}{2})
 +  4 g p\left(\frac{1}{\cos{\boldsymbol\upphi}_0}-\cos{\boldsymbol\upphi}_0\right)\,.\label{deltatotalmag}
\end{align}

\section{Strings between D5-branes at angles}
\label{obliquebranes}

In this section we will continue to study strings with large angular momentum, but introducing
a couple of modifications. Firstly, we will consider open strings stretched between two \mbox{D5-branes},
whose axis defining the AdS or S factors are oblique, {\it i.e.} at an angle $\theta$ in the S$^5$
and an angle $\phi$ in AdS$_5$. Secondly, we will consider the amount of angular momentum to be large but finite.

For such configurations, the difference $E-L$ will no longer vanish. Here we compute it
explicitly to leading order in the finite angular momentum correction. For D5-branes of the first
family, we do this in two distinct regimes: when $\tfrac{\pi}{2} - \alpha_0$ is finite and when
$\alpha_0\to \tfrac{\pi}{2}$ \footnote{An analogous distinction can be made for D5-branes of the
second family: when ${{\boldsymbol\upphi}_0}$ is finite or infinitesimal.}.
In the former, the string is long and $E-L$ can be computed classically. In the latter, the string
is short and $E-L$ has to be computed at the quantum level. This can be done because the short
string probes only the neighborhood of a null-geodesic and the lowest states in
its spectrum will be those of an open string in a pp-wave background.

The reason why these computations are useful is that the deviation of $E-L$ from $0$
can be  interpreted as a leading finite size correction which can be
independently obtained by means of a L\"uscher computation. Given that
the L\"uscher correction depends on an analytic continuation of the reflection phase, the
results of this section will therefore serve as a consistency check for an exact reflection phase proposal.

\subsection{Semiclassical string between D5-branes at angles}

We now consider a semiclassical string with large angular momentum $L$,  stretching between
two D5-branes of the first of the two families described in section \ref{classical}, when $\tfrac{\pi}{2} - \alpha_0$ is finite.
We will separate the D5-branes by an angle $\phi$ in AdS space and an angle $\theta$ in the sphere.
This computation generalizes the ones of \cite{Correa:2012hh,Gromov:2012eu} and we just focus on the large $L$ situation.

Because of the angular separation between the D5-branes, the semiclassical string propagates now
in ${\rm AdS}_3\times {\rm S}^3$. For its metric we employ coordinates
\be
ds^2 = R^2\left(\frac{dr^2}{1+r^2} - (1+r^2)dt^2 + r^2 df^2 +\frac{d\varrho^2}{1-\varrho^2} + (1-\varrho^2) d\xi_1^2 +\varrho^2 d\xi_2^2 \right)\,,
\label{ho}
\ee
and we parametrize the classical string solution as
\begin{alignat}{4}
y_1 +i y_2 &= e^{it }\sqrt{1+r^2} =e^{i\kappa \tau } \sqrt{1+r(\sigma)^2}\,,
&\qquad& & y_3 +i y_4 &= e^{i f} r =e^{i f(\sigma)} r(\sigma)\,,\label{ansatzAdS}
\\
x_1 +i x_2 &= e^{i \xi_1 } \sqrt{1-\varrho^2} =e^{i\gamma \tau } \sqrt{1-\varrho(\sigma)^2}\,,
&\qquad& & x_3 +i x_4 &= e^{i \xi_2 } \varrho = e^{i \varphi(\sigma)} \varrho(\sigma)\,.\label{ansatzS}
\end{alignat}

We work in the conformal gauge and take the range of the spatial worldsheet coordinate to be $\sigma \in [-s/2, s/2]$. The endpoints of the string are attached to D5-branes of the first family, so the boundary conditions are the ones discussed in section \ref{classical}. In the global coordinates (\ref{global})
used before, the D5-branes are placed at the azimuthal angle\footnote{Defined with respect to different oblique axes.} $\alpha_0$. When expressed in the coordinates (\ref{ho}), the position of one of the D5-branes is given by $\varrho \sin\xi_2 = \cos\alpha_0$ and the position of the other one by $\varrho \sin(\xi_2-\theta) = \cos\alpha_0$.

\begin{figure}[h]
\centering
\def\svgwidth{5.5cm}
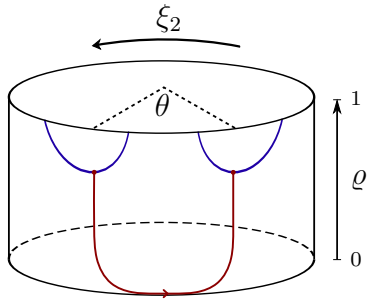
\caption{Schematically, we represent the coordinates $\varrho$ and $\xi_2$ of the metric (\ref{ho}) as a cylinder. In blue we draw the D5-branes separated by $\theta$. In red, the string between them with large angular momentum.}
\label{figD52t}
\end{figure}

In what follows we will consider a string {\it hanging} between these two D5-branes separated by an angle $\theta$, as shown in figure \ref{figD52t}.
We also contemplate the case when the D5-branes are  separated by an angle $\phi$ in a sphere within AdS.

The ansatz \eqref{ansatzAdS}-\eqref{ansatzS}, when plugged in the equations of motion and the Virasoro constraints,
leads to
\begin{alignat}{4}
 \ell_\phi &= r^2 f'\,, &\qquad\quad & &
 D_\phi & := -\ell_\phi^2 +(\kappa^2-1)r^2 +\kappa^2 r^4 = \frac{r^2 (r')^2}{1+r^2},
\label{eom1}
\\
\ell_\theta  &= \varrho^2 \varphi'\,, &\qquad\quad & &
 D_\theta & := -\ell_\theta^2 -(\gamma^2-1)\varrho^2 +\gamma^2 \varrho^4 = \frac{\varrho^2 (\varrho')^2}{1-\varrho^2},
\label{eom3}
\end{alignat}
where $D_\phi$ and $D_\theta$ are short-hand notations. The span of the spatial worldsheet coordinate can be obtained in terms of $r(\sigma)$ or $\varrho(\sigma)$ by using either \eqref{eom1} or \eqref{eom3},
\be\label{wsheetspan}
\frac{s}{2}
= \int_{r_0} ^{r_{\rm max}} \! \frac{r\ dr}{\sqrt{1+r^2}\sqrt {D_\phi}}
=\int_{\varrho_0} ^{\varrho_{\rm max}}\!\frac{\varrho\  d\varrho }{\sqrt{1-\varrho^2}\sqrt {D_\theta}}\,.
\ee
For a string with large angular momentum, $r_{\rm max} = \cot\alpha_0$ and $\varrho_{\rm max} = \cos\alpha_0$, while $r_0$ and $\varrho_0$ are the values of $r$ and $\varrho$ at $\sigma = 0$. Since we have the boundary condition $r'(0) = \varrho'(0) = 0$, they can be obtained
from
\be
0 = -\ell_\phi^2 + (\kappa^2-1)r_0^2 +\kappa^2 r_0^4\,,
\qquad
0 = -\ell_\theta^2 - (\gamma^2-1)\varrho_0^2 +\gamma^2 \varrho_0^4\,.
\label{v0u0}
\ee

Here we will just focus on a solution with $L$ very large. When the momentum $L$ and the energy $E$ go to infinity,
one has that $\varrho_0,r_0 \to 0$, that $\gamma,\kappa \to 1$ and that $\ell_\theta,\ell_\phi\to 0$.
Then, we will scale them as
\begin{align}
& \kappa = 1 + \epsilon \frac{c_\phi}{2}\,,\qquad
\ell_\phi = \epsilon \frac{\hat\ell_\phi}{2}\,,\qquad
r(\sigma) = \sqrt{\epsilon} u(\sigma)\,,
\\
& \gamma = 1 + \epsilon \frac{c_\theta}{2}\,,\qquad
\ell_\theta  = \epsilon \frac{\hat\ell_\theta}{2}\,,\qquad
\varrho(\sigma) = \sqrt{\epsilon} v(\sigma)\,.
\end{align}

The minimal values of the scaled variables become
\be\label{miniuv}
u_0^2 = \frac{-c_\phi +\sqrt{c_\phi^2+\hat\ell_\phi^2}}{2}\,,
\qquad
v_0^2 = \frac{c_\theta +\sqrt{c_\theta^2+\hat\ell_\theta^2}}{2}\,.
\ee

In the large $L$ limit, the angular span of the string is given by the angular separation
of the D-branes, {\it i.e.} $\Delta f = \pi-\phi$ and $\Delta \varphi = \theta$. By using (\ref{eom1}) and (\ref{eom3}),
the separation angles are then given in terms of $r(\sigma)$ or $\varrho(\sigma)$, and to leading order in the small $\epsilon$ expansion we have
\begin{align}
\pi-\phi & =
\int_{r_0}^{r_{\rm max}} \!\!\frac{ 2 \ell_\phi\  dr }{r\sqrt{1+r^2}\sqrt {D_\phi}}
= \int_{u_0}^\infty \frac{\hat\ell_\phi\ du }{u\sqrt{(u^2-u_0^2)(v^2+v_0^2+c_\phi)}}
= -\arctan(\hat\ell_\phi/c_\phi)\,,
\label{tan1}
\\
\theta & =
\int_{\varrho_0}^{\varrho_{\rm max}}\!\! \frac{  2\ell_\theta\ d\varrho}{\varrho \sqrt{1-\varrho^2}\sqrt {D_\theta}}
=\int_{v_0}^\infty \frac{\hat\ell_\theta\ dv }{v\sqrt{(v^2-v_0^2)(v^2+v_0^2-c_\theta)}}
= \arctan(\hat\ell_\theta/c_\theta)\,.
\label{tan2}
\end{align}

Although there is some freedom in the choice of $c_\theta$ and $c_\phi$, they are related since
the two integrals in (\ref{wsheetspan}) must agree. From the first integral, in the
small $\epsilon$ limit we get
\be
\frac{s}{2} = \log4  - \frac{1}{2}\log\left[\epsilon(2u_0^2+c_\phi)\right] -\log\left(\frac{1+\sqrt{1+r_{\rm max}^2}}{r_{\rm max}}\right)\,,
\ee
while from the second integral in (\ref{wsheetspan}) we obtain
\be
\frac{s}{2} = \log4 - \frac{1}{2}\log\left[\epsilon(2v_0^2-c_\theta)\right] -\log\left(\frac{1+\sqrt{1-\varrho_{\rm max}^2}}{\varrho_{\rm max}}\right)\,.
\ee
This implies, using \eqref{miniuv}, the relation
\be
\sqrt{c_\phi^2+\hat\ell_\phi^2},
=\sqrt{c_\theta^2+\hat\ell_\theta^2}\,.
\ee
Considering \eqref{tan1} and \eqref{tan2}, we can simply take
\be
c_\phi = \cos\phi\,,\qquad c_\theta = \cos\theta\,,
\ee
which gives\footnote{The definition of $\epsilon$ here is different than the one in \cite{Correa:2012hh}.}
\be
\epsilon = 16 e^{-s} \left(\frac{\varrho_{\rm max}}{1+\sqrt{1-\varrho_{\rm max}^2}}\right)^2\,.
\ee

As anticipated, we are interested in the difference between the energy and the angular momentum of this configuration, given by
\be
E - L = 4g\kappa \int_{r_0}^{r_{\rm max}} dr \frac{r\sqrt{1+r^2}}{\sqrt{D_\phi}} -
4g\gamma \int_{\varrho_0}^{\varrho_{\rm max}} d\varrho \frac{\varrho\sqrt{1-\varrho^2}}{\sqrt{D_\theta}}
- \left.\vphantom{\int}2 A_t\right|_{r=r_{\rm max}}\,,
\ee
where the last term comes from the boundary term due to the electric field. In the coordinates we are using
$A_t = 2 g r \cos\alpha_0$. As done is \cite{Correa:2012hh}, we
compute $L-2 g s$ and $E-2 g s$ separately. To the next to leading order in the small $\epsilon$ expansion we have
\begin{align}
\label{L-s}
L-2 g s =& -4g + g \cos\theta \epsilon  +4g\sqrt{1-\varrho_{\rm max}^2}\,,
\\
E-2 g s =&-4g + g \cos\phi \epsilon + 4g\sqrt{1+r_{\rm max}^2} - 4g r_{\rm max}\,\varrho_{\rm max}\,.
\end{align}
Given that $\sqrt{1+r_{\rm max}^2} -  r_{\rm max}\,\varrho_{\rm max} - \sqrt{1-\varrho_{\rm max}^2} = 0$,
terms which are independent of $\epsilon$ cancel in the difference, as expected. Therefore we obtain
\begin{align}
E-L &= 16 g e^{-s}\left(\frac{\varrho_{\rm max}}{1+\sqrt{1-\varrho_{\rm max}^2}}\right)^2 (\cos\phi -\cos\theta)\nn
    \\
    &= \frac{16 g}{e^{2-2\sin\alpha_0}} \tan^2\left(\tfrac{\pi}{4}-\tfrac{\alpha_0}{2}\right) (\cos\phi -\cos\theta) e^{-\frac{L}{2g}}\,,\label{EminusL_semiclassical}
\end{align}
where we used (\ref{L-s}) to express $s$ in terms of $L$.

~

For D5-branes of the second of the two families described in section \ref{classical}, the computation would follow analogously.
We do not present the details here but just the result,
\begin{align}
E-L  &= \frac{16 g}{e^{2-2\cos{\boldsymbol\upphi}_0}}\tan^2\left(\tfrac{{\boldsymbol\upphi}_0}{2}\right) (\cos\phi -\cos\theta)e^{-\frac{L}{2g}}\,.
\label{luschersinpolo2}
\end{align}

\subsection{Quantum string between oblique D5-branes  in a pp-wave}

The previous result is valid for a finite value of $\tfrac{\pi}{2} -\alpha_0$, otherwise
the semiclassical approximation is no longer valid. A string attached to a
$\alpha_0 = \tfrac{\pi}{2}$ maximal D5-brane and carrying large angular momentum will be almost point-like.
Thus, it will only probe the neighborhood of a null-geodesic, that of a particle spinning around in the $\rm{S}^5$.
Therefore, the lowest states in the string spectrum will be those of an open string in a pp-wave background with endpoints attached to a D5-brane that looks flat.

In the Penrose limit that zooms in on the null-geodesic \cite{Berenstein:2002jq,Blau:2002dy}, the metric reduces to
\be
ds^2 = -4 \, du \, dv - z^2 du^2 + d\vec{z\,}^2\,,
\ee
where $\vec z \in \mathbb{R}^8$. The D5-brane becomes flat in this limit,
sitting at $z^1 = z^2 = z^3 = 0$ for the coordinates coming from the AdS factor
and at $z^5 = 0$ for the coordinates coming from the sphere.

If we consider an open string attached to the flat brane in the pp-wave background,
given that the brane is BPS the contribution to the vacuum energy of all the bosonic and fermionic
modes of the string cancels exactly. However, we would like to consider an open string stretching between the previous D5-brane at $z^5=0$ and another one that has been rotated in the plane $(z^5, z^6)$, {\it i.e.} sitting at $\cos\theta z^5+\sin\theta z^6=0$. In other words,
the string still has Dirichlet and Neumann boundary conditions in $z^5$ and $z^6$ respectively, for the left
endpoint\footnote{The omission of a normalization factor is indicated by $\propto$.},
\begin{align}
z^5 & \propto \sum_{n} a^5_n e^{i\tau \omega_n}\sin k_n\sigma\,,
\\
z^6 &  \propto \sum_{n} a^6_n e^{i\tau \omega_n}\cos k_n\sigma\,,
\end{align}
where $\omega_n = \sqrt{m^2+k_n^2}$ for $m = \tfrac{L}{2\pi g}$.  Now in order for the string to have {\it rotated} boundary conditions
in the right endpoint, we have to take $k_n = n \mp \tfrac{\theta}{\pi}$ when $a^5_n = \pm a^6_n$.

Analogously, the fermionic modes of the string will present similar shifts, but
in their case of $ \mp \tfrac{\theta}{2\pi}$. As a consequence of all these shifts,
the vacuum energy, or more precisely  $E-L$, will no longer vanish for the open string.
We can simply compute $E-L$ as the difference between the contribution of modes with and without
the shifts,
\ba
E-L \!\!&=&\!\! \frac{1}{2m}\sum_{n=-\infty}^{\infty}\sqrt{m^2+\left(n-\tfrac{\theta}{\pi}\right)^2} -\frac{1}{2m}\sum_{n=-\infty}^{\infty}\sqrt{m^2+n^2}\nn\\
&& -\frac{2}{m}\sum_{n=-\infty}^{\infty}\sqrt{m^2+\left(n-\tfrac{\theta}{2\pi}\right)^2} +\frac{2}{m}\sum_{n=-\infty}^{\infty}\sqrt{m^2+n^2},
\ea
where the first and second lines come from bosonic and fermionic modes respectively.
It is convenient to introduce the notation
\be
h(\theta, m) := \frac{1}{m}\sum_{n=-\infty}^{\infty}\sqrt{m^2+\left(n-\tfrac{\theta}{\pi}\right)^2} -\frac{1}{m}\sum_{n=-\infty}^{\infty}\sqrt{m^2+n^2}\,,
\ee
which allows us to write
\be
E-L = \frac12 h(\theta,m)-2 h\left(\tfrac{\theta}{2}, m\right)\,.
\ee

We first consider
\be
\frac{1}{m}\partial_m (m\ h) =
\sum_{n=-\infty}^{\infty}\frac{1}{\sqrt{m^2+\left(n-\tfrac{\theta}{\pi}\right)^2}} -\sum_{n=-\infty}^{\infty}\frac{1}{\sqrt{m^2+n^2}},
\ee
and  use Poisson's resummation formula
\be
\sum_{n=-\infty}^\infty f(n) = \sum_{k=-\infty}^\infty \tilde{f}(k),
\ee
where $\tilde f$ stands for the Fourier transform of $f$\footnote{We use here the definition
\be
\tilde{f}(w) = \int_{-\infty}^{\infty} dx \, e^{-2\pi i w x} f(x) \,.\nn
\ee
}. Then
\be\label{resu}
\frac{1}{m}\partial_m (m\ h ) = \sum_{k=-\infty}^{\infty}\tilde{f}(k) \left(e^{-2 i k\theta}-1\right)\,,
\ee
for
\be
f(x) = \frac{1}{\sqrt{m^2+x^2}} \quad \Longrightarrow \quad \tilde{f}(w) = 2 K_0\left(2\pi m\left|w\right|\right)\,,
\ee
where $K_0$ is a modified Bessel function of the second kind. Now, we are only interested in the large
$m =\frac{L}{2\pi g}$ limit of this resummation. For $m$ large
\be
\tilde{f}(k) \sim \frac{e^{-2\pi \left|k\right| m}}{\sqrt{\left|k\right| m}}\,,
\ee
and the sum is dominated by $k = \pm 1$. Thus,
\be\label{resuapro}
\frac{1}{m}\partial_m (m\ h ) \sim \frac{2 e^{-2\pi m}}{\sqrt{m}}(\cos2\theta-1),
\ee
which leads to
\be
h(\theta, m) \sim -\frac{e^{-2\pi m}}{\pi \sqrt{m}}(\cos2\theta-1)\,.
\ee
Therefore
\ba
E-L \!\!&\sim&\!\! -\frac{e^{-2\pi m}}{2\pi \sqrt{m}}(\cos2\theta-1) + 2\frac{e^{-2\pi m}}{\pi \sqrt{m}}(\cos\theta-1) =
-\sqrt{\frac{2 g}{\pi L}} e^{-\frac{L}{g}} (\cos\theta-1)^2\,.
\ea

If at the same time we consider that one of the D5-branes is rotated in the plane $(z^3,z^4)$ by an angle $\phi$,
the $\omega_n$ corresponding to those bosonic coordinates will be shifted by $n\mapsto n \pm \tfrac{\phi}{\pi}$.
On the other hand, the fermionic modes' frequencies will be shifted by $n\mapsto n \pm (\tfrac{\theta}{2\pi} \pm \tfrac{\phi}{2\pi})$, so
\ba
E-L \!\!& = &\!\! \frac12 h(\phi,m)+\frac12 h(\theta,m)- h\left(\tfrac{\theta+\phi}{2},m\right) - h\left(\tfrac{\theta-\phi}{2},m\right)
\nn\\
\!\!&\sim&\!\! -\sqrt{\frac{2 g}{\pi L}} e^{-\frac{L}{g}} (\cos\theta-\cos\phi)^2\,.
\label{luscherconpolo}
\ea

An identical result is obtained for maximal D5-branes of the second family, when ${\boldsymbol\upphi}_0\to 0$.

\section{Boundary crossing condition}
\label{crossingsolutions}

The two infinite families of D5-brane boundary conditions have something in common.
All of their members preserve the same underlying symmetry: a diagonal $su(2|2)$ of the usual
$su(2|2)^2$. Then, up to an overall factor, the reflection matrices
are the same. In other words, what we ignore about the corresponding scattering problems
is restricted to an undetermined reflection factor in each case. Furthermore,
all these reflection factors are constrained by a boundary crossing condition.

For a right boundary, the undetermined reflection factor $R_0(p)$, in all the cases we consider,
must satisfy the following crossing condition \cite{Correa:2012hh,Drukker:2012de}
\be
R_0(p) R_0(\bar{p}) = \sigma(p, -\bar{p})^2\,,
\ee
where $\sigma(p_1, p_2)$ is the bulk dressing factor \cite{Beisert:2006ib,Beisert:2006ez} and $\bar{p}$ indicates a
crossing transformation, which takes a particle with energy and momentum
$(E,p)$ into a particle with energy and momentum $(-E,-p)$.
The boundary factor should also satisfy the unitarity condition
\be\label{eq:boundary_unitarity}
R_0(p) R_0(-p) = 1\,.
\ee
We will use spectral parameters $x^\pm$ to describe the kinematics of a particle, so
\be\label{eq:spectral_parameters}
\frac{x^+}{x^-} = e^{ip} \qquad \text{and} \qquad x^+ + \frac{1}{x^+} - x^- - \frac{1}{x^-} = \frac{i}{g}\,.
\ee
In terms of the spectral parameters, the crossing transformation is $x^\pm \mapsto 1/x^\pm$.

In order to deal with a simpler crossing equation, we can write the reflection factor as
\be
R_0(p) = \frac{1}{\sigma_B(p) \sigma(p, -p)} \left(\frac{1 + \tfrac{1}{\left(x^-\right)^2}}{1 + \frac{1}{\left(x^+\right)^2}}\right)\,,
\ee
where the only unknown is the boundary dressing factor $\sigma_B(p)$. Then, crossing and unitarity equations become
\be
\sigma_B(p) \sigma_B(\bar{p}) = \frac{x^- + \frac{1}{x^-}}{x^+ + \frac{1}{x^+}}\,,
\qquad
\sigma_B(p) \sigma_B(-{p}) = 1\,.
\label{crossing}
\ee

A particular boundary dressing factor that solves the system (\ref{crossing}), which we call here $\sigma^0_B(p)$,
was found in \cite{Correa:2012hh,Drukker:2012de},
\ba
  \sigma^0_B & = &  e^{ i \chi^0_B(x^+) - i \chi^0_B(x^-) }\,,\nn
\\
\label{chi0integral}
 i \chi^0_B(x) & = & i \Phi^0_B(x) =  \oint\limits_{|z|=1} { d z \over 2 \pi i } { 1 \over x - z } \log \left\{ \sinh[ 2 \pi g ( z + { 1 \over z} ) ] \over 2 \pi g ( z + { 1 \over z } ) \right\}\,,~~~~~|x|>1\,.
 \ea

In the strong coupling limit this solution reduces to the $\alpha_0\to 0$ limit of (\ref{deltatotal})
and the ${\boldsymbol\upphi}_0 \to \tfrac{\pi}{2}$ limit of (\ref{deltatotalmag}). However, to recover
(\ref{deltatotal}) and (\ref{deltatotalmag}) in the general cases we should look for new solutions of
the same crossing and unitarity conditions (\ref{crossing}). In order to do that we will take $\sigma_B(p) = \sigma_B^0(p) \sigma_T(p)$.
The unknown dressing factor $\sigma_T$ satisfies ``{\it trivial}'' crossing and unitarity conditions
\be\label{eq:trivial_crossing}
\sigma_T(p)\sigma_T(\bar{p}) = 1\,,
\qquad
\sigma_T(p) \sigma_T(-p) = 1.
\ee

As we shall see, there are infinitely many ways of solving the trivial system (\ref{eq:trivial_crossing}).
However, our analysis does not intend to be exhaustive. We will just observe that solutions obtained
in a particular way are compatible with all the strong coupling computations we have presented in the
previous sections.

We start by proposing $\sigma_T$ to be of the form
\be
\sigma_T(p)
= e^{i\chi_{T}(x^+)-i\chi_{T}(x^-)}\,,
\ee
and use a contour integral, in analogy with (\ref{chi0integral}), to define $\chi_{T}(x)$.
In particular, in terms of a generic function $F$ we define
\be
{\Phi}_F(x) = i \oint\limits_{\left|z\right| = 1} \frac{dz}{2\pi i} \frac{1}{x - z} \log F \left(z + \tfrac{1}{z}\right)\,.
\ee
We have taken the argument of the generic function to be $z + \tfrac{1}{z}$,
so that for any $F$ the contour integral satisfies
\begin{align}
\label{propertyF}
\Phi_F(x) + \Phi_F(1/x) &= \Phi_F(0)\,.
\end{align}
This property, analogue of the one discussed in \cite{Arutyunov:2009kf} for the bulk dressing phase,
will help to fulfill the boundary crossing condition. For $\chi_T(x)$ we consider solutions of the form
\be\label{eq:chi_T_general}
\chi_T(x) = \begin{cases}
\Phi_F(x) &\quad \text{if} \quad \left|x\right| > 1,\\
\Phi_F(x) + i \log F\left(x+\tfrac{1}{x}\right) &\quad \text{otherwise}.
\end{cases}
\ee

For this to give a solution of the trivial crossing condition we need
\be\label{eq:crossing_condition}
\chi_T(x^+) - \chi_T(x^-) + \chi_T(1/x^+) - \chi_T(1/x^-) = 0,
\ee
and because of property (\ref{propertyF}) this simply implies
\be
F\left(x^+ + \tfrac{1}{x^+}\right) = F\left(x^- + \tfrac{1}{x^-}\right).
\ee
Using the constraint that relates the spectral parameters $x^\pm$,
\be
F\left(x^- + \tfrac{1}{x^-}\right) = F\left(x^- + \tfrac{1}{x^-} + \tfrac{i}{g}\right)\,.
\ee
Thus, the trivial crossing equation is satisfied whenever $F$ is periodic in the imaginary axes with  period $i/g$. Concerning unitarity, it would suffice to demand that $\chi_T(x)$ is an even function, and it is straightforward to check that this is achieved whenever $F$ has definite parity, either odd or even.

Two natural possibilities are
\be
F\left(z+\tfrac{1}{z}\right) = \sinh\left[2\pi n g \left(z+\tfrac{1}{z}\right)\right]\,
\qquad{\rm or}\qquad
F\left(z+\tfrac{1}{z}\right) = \cosh\left[2\pi n g \left(z+\tfrac{1}{z}\right)\right]\,,
\label{naturalF}
\ee
for any integer $n$. However, the resulting reflection factors with these additional trivial solutions to the crossing equation
would not reproduce in general the desired strong coupling behaviors (\ref{deltatotal}) nor (\ref{deltatotalmag}). Only
the limits $\alpha_0\to\tfrac{\pi}{2}$ or ${\boldsymbol\upphi}_0 \to 0$ of (\ref{deltatotal}) and (\ref{deltatotalmag})
can be reproduced for either of these solutions when $n=1$. If we moreover wanted to use the resulting reflection factor
to obtain the finite angular momentum correction (\ref{luscherconpolo}), this would only be possible with the sinh solution.
Therefore, we would like to consider some deformations of
\be
{\Phi}_{\rm sinh}(x) = i \oint\limits_{\left|z\right| = 1} \frac{dz}{2\pi i} \frac{1}{x - z}
\log\left\{ \sinh\left[2\pi g \left(z+\tfrac{1}{z}\right)\right]\right\}\,,
\label{phisinh}
\ee
in order to obtain solutions  whose $g\to \infty $ limit is compatible with the explicit computations (\ref{deltatotal}) and (\ref{deltatotalmag}).
To make a comparison with the explicit computations of section \ref{timedelay} we will evaluate
the contribution to the reflection phase factor in the strong coupling limit due to a given solution of the trivial crossing condition as
\be
\delta_T = \chi_T(x^-)-\chi_T(x^+)\,,
\ee
for $x^\pm = e^{\pm ip/2} + {\cal O}(1/g) $. The solutions of the crossing equation should be such that
this $\delta_T$ reproduces what we called $\delta_{\rm extra}$ in section \ref{timedelay}.

\subsection{Contour integrals in re-scaled circles}\label{sec:scaledcircle}

We would like to deform somehow ${\Phi}_{\rm sinh}$ in order to introduce a dependence
with a parameter which may later be related to the amount of electromagnetic flux in the
D5-branes under consideration.

We will introduce a bold modification of ${\Phi}_{\rm sinh}$ and
check {\it a posteriori} it possess part of the desired
strong coupling dependence. In particular, the deformation
we consider in first place takes the contour of integration to be
a circle of radius $r$,
\be
{\Phi}_{r}(x) = i \oint\limits_{\left|z\right| = r} \frac{dz}{2\pi i} \frac{1}{x - z}
\log\left\{ \sinh\left[2\pi g \left(z+\tfrac{1}{z}\right)\right]\right\}\,.
\label{eq:Phi_f}
\ee
The property (\ref{propertyF}), which paves the way for solving the crossing condition,
is no longer valid for ${\Phi}_{r}$. There is, nevertheless, a useful deformation
of it,
\begin{align}
\label{propertydef}
\Phi_r(x) + \Phi_{1/r}(\tfrac{1}{x}) &= \Phi_r(0)\,.
\end{align}
Then, by combining two contour integrals of sizes $r$ and $\tfrac{1}{r}$ in
\be\label{eq:Phi_T}
\Phi_T(x) = \frac{1}{2}\left(\Phi_r(x) + \Phi_{1/r}(x)\right)\,,
\ee
we obtain a function with the desired property
\be\label{eq:Phi_T_x_Phi_T_1/x}
\Phi_T(x) + \Phi_T\left(\tfrac{1}{x}\right) = \Phi_T(0)\,.
\ee
For definiteness we take $0<r\leq1$, and use the combination $\Phi_T(x)$ to define $\chi_T(x)$ in the region outside the
circle of radius $1/r$. From this region, we can analytically continue to anywhere in the plane to have
\be\label{eq:chi_T}
\chi_T^{(1)}(x) = \begin{cases}\Phi_T(x) \hfill \left|x\right| > 1/r\,,&\\
\Phi_T(x) +\frac{i}{2} \log \sinh 2\pi g \left(x + \tfrac{1}{x}\right) \qquad\qquad  \hfill r < \left|x\right| < 1/r\,, &\\
\Phi_T(x) + i \log \sinh 2\pi g \left(x + \tfrac{1}{x}\right)  \hfill \left|x\right| < r\,. \,\,\,\,\,\,&
\end{cases}
\ee
If we now used relation \eqref{eq:Phi_T_x_Phi_T_1/x} as before, we could check explicitly the trivial crossing condition
\eqref{eq:crossing_condition} is satisfied.

The solution to the trivial crossing equation from this $\chi_T^{(1)}$ is valid for all values of the coupling $g$.
However, when considered in the strong coupling limit, as we will see in what follows, it can only explain one of the
extra terms in the boundary reflection  phase (\ref{delta41}). We have added $^{(1)}$ to indicate that.

Now, we want to describe what the resulting reflection phase factor would be for
particles with physical kinematics, {\it i.e.} with $|x^\pm |>1$, in the strong coupling limit.
For $\left|x\right| > r$ we can expand $(x-z)^{-1}$ as a geometric series in our definition of $\Phi_r(x)$, in order to get
\be
\Phi_r(x) = \frac{1}{2\pi i} \sum_{n=1}^\infty \frac{c_n(r)}{x^n} \qquad \text{with} \qquad c_n(r) = i \oint\limits_{\left|z\right| = r} dz \, z^{n-1} \log\left\{\sinh\left[2\pi g\left(z + \tfrac{1}{z}\right)\right]\right\}.
\ee
The imaginary part of the coefficients $c_n(r)$ can be seen to vanish, as well as the real part whenever $n$ is odd. The remaining coefficients can be evaluated in the strong coupling limit $g \to \infty$, giving
\be
c_{2k}(r) = -8\pi g (-1)^k r^{2k} \left(\frac{r}{1+2k} + \frac{r^{-1}}{1-2k}\right),
\ee
which upon resummation leads to
\be\label{eq:largeg_Phif}
\Phi_r(x) = 4ig \left(x + \frac{1}{x}\right)\arctan\left(\frac{r}{x}\right) + {\cal O}(g^0),
\ee
up to $x$-independent terms that will cancel when we compute the reflection phase factor. Using the definitions of $\Phi_T(x)$ and $\chi_T(x)$, as well as property \eqref{propertydef}, we can now evaluate the reflection phase factor in the strong coupling limit due to this trivial crossing solution as
\be
\delta_T^{(1)} = -4g \cos \tfrac{p}{2} \log \frac{r^2 + 1 + 2 r \sin\tfrac{p}{2}}{r^2 + 1 - 2 r \sin\tfrac{p}{2}}\,.
\ee
Now, if we take
\be\label{eq:r}
r = \cot(\tfrac{{\boldsymbol\upphi}_0}{2}+\tfrac{\pi}{4})\,,
\ee
we obtain
\be
\delta_T^{(1)} = -4g \cos\tfrac{p}{2} \log \frac{ 1 + \cos{\boldsymbol\upphi}_0\sin\tfrac{p}{2}}{ 1 - \cos{\boldsymbol\upphi}_0\sin\tfrac{p}{2}}\,.
\ee
This is only one of the extra terms in the boundary reflection phase (\ref{delta41}), more precisely the first one.
The other terms also need to be explained in terms of solutions to the trivial crossing and unitarity conditions.
For instance, if we consider
\be
\chi_{T}^{(2)}(x) = f_2\left({\boldsymbol\upphi}_0, g\right) \log\frac{x\,r + 1/x\,r}{x/r + r/x}\,,
\label{cacho2}
\ee
with $r$ defined in \eqref{eq:r}, we see that the resulting $\sigma_T^{(2)}(p)$ satisfies the trivial crossing equation as well as the unitarity condition, while contributing in the strong coupling limit to the reflection phase factor in
\be
\delta^{(2)}_T = 2 f_2\left({\boldsymbol\upphi}_0, g\right) \log\frac{i - \sin{\boldsymbol\upphi}_0 \tan \tfrac{p}{2}}{i + \sin{\boldsymbol\upphi}_0 \tan\tfrac{p}{2}} = 4 i f_2\left({\boldsymbol\upphi}_0, g\right) \arctan(\sin{\boldsymbol\upphi}_0 \tan\tfrac{p}{2})\,.
\ee
This corresponds to the second extra term in the reflection phase factor calculated in section \ref{subsec:delta_ads4s2} whenever $f_2\left({\boldsymbol\upphi}_0, g\right)$ behaves like $2gi \tan {\boldsymbol\upphi}_0$ in the strong coupling limit.

The same can be done to take into account the third extra term in the reflection phase factor, by simply taking
\be
\chi_{T}^{(3)}(x) = f_3({\boldsymbol\upphi}_0, g) \log x\,,
\label{cacho3}
\ee
with a suitable $g\to\infty$ limit for $f_3({\boldsymbol\upphi}_0, g)$, namely $f_3({\boldsymbol\upphi}_0, g) \sim 4 g i \left(\displaystyle\frac{1}{\cos{\boldsymbol\upphi}_0} - \cos{\boldsymbol\upphi}_0\right)$.

\subsection{Line integrals in arcs}\label{sec:arc}

We will now consider another way in which we can modify our initial proposal $\Phi_{\sinh}$. We shall consider
\be\label{eq:Phi_arcos}
\Phi_\gamma(x) = i \int\limits_{\mathcal{C}(\gamma)} \frac{dz}{2\pi i} \frac{1}{x-z} \log \left\{\sinh \left[2 \pi g \left(z + \tfrac{1}{z}\right)\right]\right\},
\ee
for an open curve $\mathcal{C}(\gamma)$ parameterized by $z(t) = e^{i t}$ with $-\gamma < t < \gamma$ and $\pi-\gamma < t < \pi+\gamma$. Because this curve is
invariant under $z \mapsto 1/z$ and we use a function of $z + \tfrac{1}{z}$ only, property \eqref{propertyF} will hold. Being defined as an integration along an
open curve, we may propose that $\chi_T$ is defined by just the line integral outside as well as inside the unit disk.
Then, property \eqref{propertyF} would suffice to conclude that \eqref{eq:Phi_arcos} solves the trivial crossing equation.
It is not, however, an even function of $x$, so in order to satisfy the unitarity condition we can define $\chi_T(x) = \tfrac{1}{2}(\Phi_\gamma(x)+\Phi_\gamma(-x))$.

A strong coupling analysis of this proposal for  $|x|>1$ can be performed along the lines of the previous section, and we obtain
\be
\chi_T^{(1)}(x) = \frac{1}{2\pi i} \sum_{k=1}^\infty \frac{c_{2k}(\gamma)}{x^{2k}}
\quad{\rm with}\quad
c_{2k}(\gamma) = -16 \pi g \frac{2k \cos\gamma \sin(2k\gamma) - \sin \gamma \cos(2k\gamma)}{4k^2 - 1} 
\ee
up to subleading terms in the limit $g \to \infty$. This gives, up to $x$-independent terms that will cancel when we compute the reflection phase factor,
\be
\chi_T^{(1)}(x) = 2g \left(x + \frac{1}{x}\right) \left({\rm arctanh}\frac{e^{i\gamma}}{x} - {\rm arctanh}\frac{e^{-i\gamma}}{x}\right)+ {\cal O}(g^0).
\label{chi1arc}
\ee
As before, the $^{(1)}$ indicates this solution to the trivial crossing condition would explain only
the first term in \eqref{delta1}. To see this we evaluate the reflection phase factor corresponding
to this solution of the crossing equation,
\be
\delta_T^{(1)} = -4g \cos \tfrac{p}{2} \log\left|\frac{\sin\tfrac{p}{2} + \sin\gamma}{\sin\tfrac{p}{2} - \sin\gamma}\right|\,.
\label{deltaT1}
\ee
If we identify $\gamma=\alpha_0$, this is the first extra term in the boundary reflection phase \eqref{deltatotal}.
In order to explain the second term we can propose
\be
\chi_T^{(2)}(x) = f(\alpha_0,g)\log\left|\frac{e^{-i\alpha_0}x- \tfrac{e^{i\alpha_0}}{x}}{e^{i\alpha_0}x- \tfrac{e^{-i\alpha_0}}{x}}\right|\,,
\label{chiT2}
\ee
which is a solution of the trivial crossing and unitarity conditions and leads to
\be
\delta_T^{(2)} = 2f(\alpha_0,g) \log\left|\frac{\sin(\tfrac{p}{2} +\alpha_0)}{\sin(\tfrac{p}{2} -\alpha_0)}\right|\,.
\label{deltaT2}
\ee
This would be the second extra term in \eqref{deltatotal} for any $f(\alpha_0,g)$ whose strong coupling limit is $2 g\cos\alpha_0$.

As remarked before, terms (\ref{deltaT1}) and (\ref{deltaT2}) of the resulting reflection phase become
logarithmically divergent as $p \to \pm 2\alpha_0$, but they cancel for the proposed limiting value of $f(\alpha_0,g)$.
In the exact $\chi_T^{(1)}$ and $\chi_T^{(2)}$ proposals these correspond to the logarithmic divergencies
when $x \to \pm e^{\pm \alpha_0}$. In particular, the logarithmic divergence of $\chi_T^{(1)}(x)$  appears when $x$ is
evaluated at the endpoints of $\mathcal{C}(\alpha_0)$.

The choice $f(\alpha_0,g) =  g\cos\alpha_0 +{\cal O}(g^0)$, necessary to match the explicit strong coupling computation,
ensures nevertheless that $\chi_T$ is regular in that limit. If we require that $\chi_T$ continues to be regular at
$x = \pm e^{\pm \alpha_0}$ to all orders in $1/g$, this would allow us to determine $f(\alpha_0, g)$ exactly.
In order to do this, we first observe that the proposal \eqref{chiT2} can also take the form of an integration along
$\mathcal{C}(\alpha_0)$, namely
\be
\chi_T^{(2)}(x) = -f(\alpha_0, g) \int\limits_{\mathcal{C}(\alpha_0)} \frac{dz}{x - z}\,,
\ee
up to $x$-independent terms that will cancel when we compute $\sigma_T$. Then we can write $\chi_T$ as a single integration along $\mathcal{C}(\alpha_0)$.
For the integral to be regular for  $x = \pm e^{\pm \alpha_0}$, we should demand
that the factor accompanying $\left(x-z\right)^{-1}$ vanishes as $z$ approaches the endpoints of the curve. This fixes $f(\alpha_0, g)$ to be
\be
2\pi f(\alpha_0, g) = \log \left[\sinh\left(4\pi g \cos \alpha_0\right)\right]\,,
\ee
which in turn means that we can write
\be
\chi_T(x) = i \int\limits_{\mathcal{C}(\alpha_0)} \frac{dz}{4\pi i} \left(\frac{1}{x - z} + \frac{1}{-x - z}\right)\log\left[ \frac{\sinh\left( 2\pi g \left(z + \tfrac{1}{z}\right) \right)}{\sinh\left( 4\pi g \cos\alpha_0\right)} \right]\,.
\ee

~

Note that this deformation, just as the one presented in the previous section when $r \to 1$, reduces to $\Phi_{\sinh}$ when $\alpha_0 \to \tfrac{\pi}{2}$\footnote{When $\alpha_0 \to \tfrac{\pi}{2}$ the curve $\mathcal{C}(\alpha_0)$ closes to form the unit circle and the contribution of $\chi_T^{(2)}$ vanishes.}. In this limit it is in fact convenient to consider the full quantities
\be
\sigma_B(p) = \sigma_B^0(p) \, \sigma_T(p) = e^{i \chi_B(x^+) - i \chi_B(x^-)} \qquad \text{with} \qquad \chi_B(x) = \chi_B^0(x) + \chi_T(x)\,,
\label{dege1}
\ee
where
\be
\Phi_B(x) = i \oint\limits_{\left|z\right| = 1} \frac{dz}{2\pi i} \frac{\log\left[2\pi g \left(z + \tfrac{1}{z}\right)\right]}{x - z}\,,
\label{dege2}
\ee
and $\chi_B(x) = \Phi_B(x)$ for $\left|x\right| > 1$, and an additional term $i\log\left[2\pi g\left(x + \tfrac{1}{x}\right)\right]$ should be added to ensure continuity if $\left|x\right| < 1$. As expected, the contribution to the reflection phase $\delta$ from this $\sigma_B$
is order $g^0$ rather than order $g$.

\subsection{Further verifications from L\"uscher computations}
\label{luschersec}

So far we have compared the strong coupling limit of some solutions to the crossing equation with explicit
computations of the boundary reflection phase factors performed in section \ref{timedelay}.

We can also use the results of section \ref{obliquebranes} to further test compatible
solutions of the crossing equation. Computations of section \ref{obliquebranes} should be interpreted
as leading finite angular momentum corrections to the value of $E-L$ for open strings between D5-branes at angles.
Then, we should be able to reproduce those results by a L\"uscher computation. Since L\"uscher
computations depend on an analytic continuation of the  boundary reflection phase factors,
we can use this to further restrict which solutions to the crossing equation are admissible for the reflection phase factors.

The boundary L\"uscher correction \cite{LeClair:1995uf} can be obtained from
\be
E-L \sim - \sum_{a=1}^\infty\int_0^\infty \frac{dq}{2\pi} \log
\left[1+e^{-2L\tilde{E}_a(q)}t_a(q)\right]\,,
\label{luscha}
\ee
where
\be
t_a(q) = \sigma_B \bar\sigma_B \left(\frac{z^{[-a]}}{z^{[+a]}}\right)^2
\left[2(-1)^a(\cos\phi-\cos\theta)\frac{\sin a\phi}{\sin\phi}\right]^2\,,
\ee
and
\be
\tilde{E}_a(q) = 2\ {\rm arcsinh}\left(\frac{\sqrt{a^2+q^2}}{4g}\right) \,.
\ee
In the equation for $t_a(q)$, $\sigma_B $ and $\bar\sigma_B$ are shorthand notations for
\be
\sigma_B := \sigma_B\left(z^{[+a]},z^{[-a]}\right)\,\qquad  \bar\sigma_B := \sigma_B\left(-\frac{1}{z^{[-a]}},-\frac{1}{z^{[+a]}}\right)\,,
\ee
and
\be
z^{[\pm a]} = \frac{q+ia}{4g}\left(\sqrt{1+\frac{16g^2}{a^2+q^2}}\pm 1\right)
\ee
are the spectral parameters of particles with mirror kinematics. Therefore, the L\"uscher computation
in the present cases will be almost the same as the one discussed in \cite{Correa:2012hh,Drukker:2012de},
except for a different boundary phase factor $\sigma_B$.

We will evaluate (\ref{luscha}) in the limit $1 \ll g \ll L$, where $\tilde{E}_a(q) \sim \frac{\sqrt{a^2 + q^2}}{2 g}$, so that the integration will be dominated by the region $q \ll 1$, and the sum over $a$ will therefore always be dominated by the $a = 1$ term. If, as in the case of the fundamental Wilson loop, the quantity $\sigma_B \bar\sigma_B$ had a double pole, the leading L\"uscher correction would be dominated by a single mirror-particle exchange \cite{Bajnok:2004tq}, as in the case discussed in \cite{Correa:2012hh,Drukker:2012de},
\be
E-L \sim -\frac{1}{2} e^{-\frac{L}{2g}} \sqrt{\left.\left(q^2 t_1(q)\right)\right|_{q=0}}\,.
\ee
On the other hand, if $\sigma_B \bar\sigma_B$ goes to a constant as $q\to 0$, the leading L\"uscher correction
would be dominated by a pair of mirror-particles exchange and
\be
E-L \sim -\int_0^{\infty} \frac{dq}{2\pi} e^{-2L\tilde{E}_1(q)} t_1(q)
\sim -\frac{t_1(0)}{4} e^{-\frac{L}{g}} \sqrt{\frac{2g}{L \pi}}\,.
\label{Luscherpairexchange}
\ee

Recall that we have written $\sigma_B = \sigma_B^0 \sigma_T$ where $\sigma_B^0$ is the
fundamental representation boundary dressing factor, which has a pole, and
$\sigma_T$ is a solution of the trivial crossing equation. Then, depending on how $\sigma_T \bar\sigma_T$ behaves
as $q$ goes to zero, we can face any of the two possibilities mentioned above:
if $\sigma_T \bar\sigma_T$ goes to a non-vanishing constant $\sigma_B \bar\sigma_B$ continues to have
a double pole; on the other hand, if $\sigma_T \bar\sigma_T$ has a double zero as $q \to 0$, this shall cancel the double pole in $\sigma_B^0 \bar\sigma_B^0$ and leave us with a regular $\sigma_B \bar\sigma_B$ at $q = 0$.

A glance at the leading finite angular corrections (\ref{EminusL_semiclassical})-(\ref{luschersinpolo2}) and
(\ref{luscherconpolo}) leads us to expect that $\sigma_B \bar\sigma_B$ should have a double pole in the
general case but become regular as $\alpha_0\to\tfrac{\pi}{2}$ or ${\boldsymbol\upphi}_0\to0$.

To analyze this, we evaluate $\sigma_T \bar\sigma_T$ defined in terms of the $\chi_T$ introduced in sections \ref{sec:scaledcircle} and
 \ref{sec:arc}, as $q$ goes to $0$ and in the limit $g\to \infty$, that is, we want to evaluate
\be
\sigma_T \bar\sigma_T = e^{i\left[\chi_T(z^{[+a]}) - \chi_T(z^{[-a]}) + \chi_T(-1/z^{[-a]}) - \chi_T(-1/z^{[+a]})\right]}\,.
\ee

Let us consider first the $\chi_T$ obtained in section \ref{sec:scaledcircle}, which is the sum of
the solutions to the trivial crossing equation (\ref{eq:chi_T}), (\ref{cacho2}) and (\ref{cacho3}),
$\chi_T = \chi_T^{(1)} + \chi_T^{(2)} + \chi_T^{(3)}$. Because $z^{[\pm a]} \to i$ as $q\to 0$,  we
have to use $r < \left| z^{[\pm a]}\right| < 1/r$ in the definition (\ref{eq:chi_T}). The case $r = 1$,
{\it i.e.} ${\boldsymbol\upphi}_0 = 0$, degenerates and will be studied separately.

The extra terms in the definition (\ref{eq:chi_T}) cancel exactly, so we only need to deal with contour integrals
$\Phi_T$. Then if we use property \eqref{propertydef}, we get
\be
\left.\sigma_T \bar\sigma_T \right|_{q = 0}  =
\left.e^{i\left[\Phi_r(z^{[+a]}) - \Phi_r(1/z^{[+a]}) + \Phi_r(1/z^{[-a]}) - \Phi_r(z^{[-a]})\right]}
\sigma_T^{(2)} \bar\sigma_T ^{(2)} \sigma_T^{(3)} \bar\sigma_T ^{(3)}\right|_{q = 0}\,.
\ee
For all the terms, the argument of $\Phi_r$ has norm greater than $r$ in the $q \to 0$ limit, so we can use \eqref{eq:largeg_Phif}
to evaluate them in the strong coupling limit. For $0 < {\boldsymbol\upphi}_0 < \frac{\pi}{2}$ we see that $\sigma_T \bar\sigma_T$ is regular at $q = 0$ in this limit, giving us
\be
\left.\sigma_T \bar\sigma_T\right|_{q = 0} \sim \tan\left(\frac{{\boldsymbol\upphi}_0}{2}\right)^{4a} e^{4a\cos{\boldsymbol\upphi}_0}\,.
\ee
Thus, for non-vanishing ${\boldsymbol\upphi}_0$, $\sigma_B \bar\sigma_B$ continues to have the double pole. Then the L\"uscher result in the limit $1 \ll g \ll L$
will acquire an extra factor of $\tan\left(\frac{{\boldsymbol\upphi}_0}{2}\right)^{2} e^{2\cos{\boldsymbol\upphi}_0}$ in comparison with that
of \cite{Correa:2012hh}, leading to
\be
E - L \sim \frac{16 g}{e^{2-2\cos\boldsymbol\upphi_0}} \left(\cos\phi - \cos\theta\right) \tan^2 \left(\tfrac{{\boldsymbol\upphi}_0}{2}\right) e^{-\frac{L}{2g}}\,,
\ee
which is exactly (\ref{luschersinpolo2}).

~

We now turn to a similar L\"uscher computation, but employing this time $\chi_T = \chi_T^{(1)} + \chi_T^{(2)}$ presented in section \ref{sec:arc}. Let us focus for the moment on the contribution coming from the line integral. In this case, since
\be\label{zeta0}
z_0 :=\left.z^{[+a]}\right|_{q=0} =-\left.\frac{1}{z^{[-a]}}\right|_{q=0},
\ee
approaches to $i$ in the large $g$ limit, we cannot just use (\ref{chi1arc}) for the L\"uscher computation.
Instead, we consider
\be
\left.\sigma_T^{(1)} \bar\sigma_T^{(1)} \right|_{q = 0} = e^{i\left[\Phi_\gamma(z_0) - \Phi_\gamma(1/z_0) + \Phi_\gamma(-z_0) - \Phi_\gamma(-1/z_0)\right]}\,,
\ee
with
\begin{align}
\log\left(\left.\sigma_T^{(1)} \bar\sigma_T^{(1)} \right|_{q = 0}\right) & = \int\limits_{\mathcal{C}(\gamma)} \frac{dz}{2\pi i}
\frac{2 z (1-z_0^4)}{(z^2-z_0^2)(1-z^2 z_0^2 )}\log \left\{\sinh \left[2 \pi g \left(z + \tfrac{1}{z}\right)\right]\right\}.
\end{align}
In this expression it is safe to take the large $g$ limit before integrating. We get
\begin{align}
\log\left(\left.\sigma_T^{(1)} \bar\sigma_T^{(1)} \right|_{q = 0}\right) & = -4a\int\limits_{0}^\gamma \frac{1}{\cos t} +{\cal O}(1/g)\nn\\
& = 4a\log\left[\tan\left(\tfrac{\pi}{4}-\tfrac{\gamma}{2}\right)\right]  +{\cal O}(1/g).
\end{align}
We should recall that $\gamma=\alpha_0$ in this case. The contribution from $\chi_T^{(2)}$ can be directly
evaluated using its definition (\ref{chiT2}). The total contribution of $\sigma_T$ is
\be
\left.\sigma_T \bar\sigma_T\right|_{q = 0} \sim \tan\left(\frac{\pi}{4}-\frac{\alpha_0}{2}\right)^{4a} e^{4a\sin\alpha_0}\,,
\ee
and the full L\"uscher result in the limit $1 \ll g \ll L$ will be
\be
E - L \sim \frac{16 g}{e^{2-2\sin\alpha_0}} \left(\cos\phi - \cos\theta\right) \tan^2 \left(\tfrac{\pi}{4}-\tfrac{\alpha_0}{2}\right) e^{-\frac{L}{2g}}\,,
\ee
which coincides with the explicit computation (\ref{EminusL_semiclassical}).

Let us conclude this section with the L\"uscher computation in the cases ${\boldsymbol\upphi}_0 = 0$ and $\alpha_0=\tfrac{\pi}{2}$.
For both of them the proposed boundary dressing factor is given by \eqref{dege1}-\eqref{dege2}. As we anticipated, for the L\"uscher
computation to agree with the explicit result \eqref{luscherconpolo}, $\sigma_B \bar\sigma_B$ has to be regular at $q=0$
\footnote{At this point it becomes evident that the cosh solution of \eqref{naturalF} is not suitable.
The additional term in the definition of $\chi_B$ for $|x|<1$ would in this case be $i\log\left\{2\pi g\left(x + \tfrac{1}{x}\right)\cot \left[2\pi g\left(x + \tfrac{1}{x}\right)\right]\right\}$ and $\sigma_B \bar\sigma_B$ would continue to have a double pole.}.
Then,
\be
E-L \sim -\frac{t_1(0)}{4} e^{-\frac{L}{g}} \sqrt{\frac{2g}{L \pi}}
    \sim -\sqrt{\frac{2 g}{\pi L}} e^{-\frac{L}{g}} (\cos\theta-\cos\phi)^2 \left. \sigma_B \bar\sigma_B\right|_{q=0}\,,
\ee
which would agree with \eqref{luscherconpolo} provided $\left. \sigma_B \bar\sigma_B\right|_{q=0}=1$.
To see this is indeed the case, we write
\be
\left.\sigma_B \bar\sigma_B \right|_{q = 0} = e^{2i\left[\chi_B(z_0) - \chi_B(-1/z_0)\right]} = 4 \pi^2 g^2 \left(z_0 + \frac{1}{z_0}\right)^2 e^{2i \left[\Phi_B(z_0) - \Phi_B(-1/z_0)\right]},
\label{sigmasigma}
\ee
where $z_0$ is as defined in \eqref{zeta0}. We have
\begin{align}
\Phi_B(z_0) - \Phi_B\left(-1/z_0\right) &= i \oint\limits_{\left|z\right| = 1} \frac{dz}{2\pi i} \frac{z_0^2 + 1}{\left(z_0 - z\right)\left(z_0 \, z + 1\right)} \log\left[2\pi g\left(z + \tfrac{1}{z}\right)\right]\\
&= \frac{2i}{\pi}\left(z_0^4 - 1\right) \int_0^{\pi/2} \frac{\log \left(4\pi i g \sin t\right)}{\left(z_0^2 + 1\right)^2 - 4 z_0^2 \sin^2 t} dt.
\end{align}
This time, we should do the integral before considering the large $g$ limit.
The result of the integral is quite complicated, but at the end of the day we get
\begin{align}
e^{2i \left[\Phi_B(z_0) - \Phi_B(-1/z_0)\right]} &= -\frac{1}{a^2 \pi^2} + \mathcal{O}(1/g)\,,
\end{align}
which, altogether with the other factor in \eqref{sigmasigma}, leads to $\left.\sigma_B \bar\sigma_B\right|_{q=0} = 1$ as expected.

\section{Conclusions}
\label{conclusions}

We have studied the scattering problem for excitations along open strings ending on certain D5-branes.
We have considered two kinds of D5-branes: with worldvolume AdS$_2\times$S$^4$ and some electric field
and with worldvolume AdS$_4\times$S$^2$ and some magnetic field. The D5-branes of the first type are the
dual description of $\tfrac{1}{2}$-BPS Wilson loops in the $k$-th rank antisymmetric representation of the
SU($N$) in ${\cal N}=4$ super Yang-Mills theory. The D5-branes of the second type provide the dual description
of a conformal field theory with fundamental hypermultiplets on a 2+1-dimensional defect, with some of the fundamental fields having a vacuum expectation value.

The exact determination of reflection matrices would allow in one case the exact computation of expectation values of
deformations of the antisymmetric representation $\tfrac{1}{2}$-BPS Wilson loops, by insertions of composite
operators in the adjoint representation. In the other case it would allow the
exact description of the spectral problem in the defect conformal field theory. The underlying symmetry, in both cases a diagonal $su(2|2)$ of the usual
$su(2|2)^2$, fixes the matrix structure, the asymptotic nested Bethe equations and the thermodynamic Bethe ansatz
system \cite{Correa:2012hh,Drukker:2012de,Correa:2008av}. All this is up to a reflection phase factor $\sigma_B$,
which is a function of the momentum of the reflected particles and the coupling constant $g$.

In this article we have precisely studied $\sigma_B$ for the D5-branes mentioned above.
In first place, we have explicitly computed $\sigma_B$ in the strong coupling limit, by relating
it to the time delay of reflected worldsheet solitons. For the two cases under study these explicit results
can be found in subsections \ref{subsec:delta_ads2s4} and \ref{subsec:delta_ads4s2}. We proceeded in section
\ref{obliquebranes} with the explicit computation of $E-L$ to leading order in $L$ large but finite in the strong
coupling limit, for open strings between D5-branes at angles. These  are also useful results given that, by means of a
L\"uscher computation, they can be related to certain analytic continuation of $\sigma_B$.

Finally, in section \ref{crossingsolutions} we have studied solutions to the crossing and unitarity conditions
that all the reflection factors $\sigma_B$ must satisfy. There are infinitely many solutions to these equations.
However, we have singled out some solutions consistent  with all the explicit computations of sections \ref{timedelay} and
\ref{obliquebranes}. The boundary reflection factor can always be written as
\be
R_0(p) = \frac{1}{\sigma_B^0(p)\sigma_T(p) \sigma(p, -p)} \left(\frac{1 + \tfrac{1}{\left(x^-\right)^2}}{1 + \frac{1}{\left(x^+\right)^2}}\right)\,,
\ee
where $\sigma_B^0$ is the boundary dressing phase (\ref{chi0integral}) proposed in \cite{Correa:2012hh,Drukker:2012de} and
$\sigma_T = e^{i\chi_T(x^+)-i\chi_T(x^-)}$ is an extra boundary dressing factor that solves the system (\ref{eq:trivial_crossing}).
For the D5-branes of the first family, dual to $\tfrac{1}{2}$-BPS Wilson loops in antisymmetric representations,  we propose
\be
\chi_T(x) = i \int\limits_{\mathcal{C}(\alpha_0)} \frac{dz}{4\pi i} \left(\frac{1}{x-z} + \frac{1}{-x-z}\right) \log\left[\frac{\sinh\left(2\pi g \left(z + \tfrac{1}{z}\right)\right)}{\sinh\left(4\pi g \cos\alpha_0\right)}\right]\,,
\label{propo1}
\ee
with $\mathcal{C}(\alpha_0)$ parameterized by $z(t) = e^{i t}$ with $-\alpha_0 < t < \alpha_0$ and $\pi-\alpha_0 < t < \pi+\alpha_0$.
In the proposal for this case the dependence on the coupling constant $g$ is fully fixed.

On the other hand, for D5-branes of the second family we have
\begin{align}
\chi_T(x) &= i \oint\limits_{\left|z\right| = r} \frac{dz}{4\pi i} \frac{\log\left\{\sinh\left[2\pi g \left(z + \tfrac{1}{z}\right)\right]\right\}}{x - z} + i \oint\limits_{\left|z\right| = 1/r} \frac{dz}{4\pi i} \frac{\log\left\{\sinh\left[2\pi g \left(z + \tfrac{1}{z}\right)\right]\right\}}{x - z}\nn\\
&\quad+ f_2({\boldsymbol\upphi}_0, g) \log\frac{xr + 1/xr}{x/r + r/x} + f_3({\boldsymbol\upphi}_0, g) \log x
\label{propo2}
\end{align}
for $\left|x\right| > 1/r$, and additional terms should be added as in \eqref{eq:chi_T} when $|x| < 1/r$.
In this proposal, some functions of the coupling are only determined to leading order in $g$, since we lack an argument similar
to the one that allowed us to fix the corresponding undetermined function in the previous case.

For the boundary dressing factors proposed here we had to distinguish between two regimes: when $\alpha_0$ or ${\boldsymbol\upphi}_0$ take generic values and
(\ref{propo1}) and (\ref{propo2}) are valid, and when $\alpha_0=\tfrac{\pi}{2}$ or ${\boldsymbol\upphi}_0 = 0$ (and the spherical factors of the D5-branes are maximal). For the latter cases, our proposals become
\be
\chi_T(x) = i \oint\limits_{\left|z\right| = 1} \frac{dz}{2\pi i} \frac{\log\left\{\sinh\left[2\pi g \left(z + \tfrac{1}{z}\right)\right]\right\}}{x - z} \qquad \text{for} \qquad \left|x\right| > 1\,.
\ee
which cancels an identical term in $\chi_B^0$ and the complete dressing is given in terms of
\be
\chi_B(x) = i \oint\limits_{\left|z\right| = 1} \frac{dz}{2\pi i} \frac{\log\left[2\pi g \left(z + \tfrac{1}{z}\right)\right]}{x - z} \qquad \text{for} \qquad \left|x\right| > 1\,.
\ee

This solution to the crossing and unitarity conditions for maximal D5-branes is also determined
for all values of the coupling. In this case, the verification of the L\"uscher computation
only required that, for mirror kinematics, $\left. \sigma_B \bar\sigma_B\right|_{q=0}=1$.
The Poisson resummation required to compute $E-L$ for the ground state of an open string in a
pp-wave is essentially the same as the L\"uscher computation \eqref{Luscherpairexchange} when the corrections
are dominated by the exchange of a pair of mirror particles. In first place the L\"uscher formula
\eqref{luscha} is derived by treating the exchanged particles between the boundary states as free.
Then, the fact that the Poisson sum is dominated by the terms with $k=\pm 1$, \eqref{resuapro}, is the same
as the L\"uscher computation \eqref{luscha} being dominated by a pair of mirror particles exchange.

An interesting aspect of the L\"uscher computations in these two regimes is that
for generic D5-branes the leading finite angular momentum correction is order $e^{-{L}/{2g}}$, while
for maximal D5-branes it is order $e^{-L/g}$. This was understood in terms of the proposed
boundary dressing factors $\sigma_B$ which degenerate in the maximal D5-brane limit and no longer possess the pole
that explained the order $e^{-{L}/{2g}}$ in the generic case.

A natural direction for a future work complementing our results would be to study the
dressing factors $\sigma_B$ in the weak coupling limit. This would provide more verifications
and could shed more light on the undetermined functions in (\ref{propo2}).

~

\noindent
{\bf Acknowledgements }

We would like to thank B.Basso, J.Maldacena and A.Sever for discussions.
The research of D.H.C. is supported by CONICET and grant PICT 2010-0724.
The research of F.I.S. is supported by CONICET.


\end{document}